\pdfoutput=1

\documentclass[aps,prl,reprint,twocolumn,preprintnumbers,superscriptaddress,letterpaper]{revtex4-2}
\usepackage[english]{babel}
\usepackage{ucs}
\usepackage[utf8x]{inputenc}
\usepackage{amsmath}
\usepackage{amsfonts}
\usepackage{amssymb}
\usepackage{graphicx}
\usepackage{bm}
\usepackage{bbm}
\usepackage{empheq}
\usepackage{xr-hyper}
\usepackage{hyperref}
\usepackage{xcolor}

\newcommand{\ea}{\emph{et.\,al.}}

\begin{document}
\title{Saving superconducting quantum processors from qubit decay \\ and correlated errors generated by gamma and cosmic rays}
\author{John M. Martinis}
\affiliation{Quantala, Santa Barbara, CA 93105, USA}
\email{martinis@quantala.tech}
\date{\today}

\begin{abstract}
Error-corrected quantum computers can only work if errors are small and uncorrelated.  Here I show how cosmic rays or stray background radiation affects superconducting qubits by modeling the phonon to electron/quasiparticle down-conversion physics.  For present designs, the model predicts about 57\% of the radiation energy breaks Cooper pairs into quasiparticles, which then vigorously suppress the qubit energy relaxation time ($T_1 \sim 600\,$ns) over a large area (cm) and for a long time (ms).  Such large and correlated decay kills error correction.   Using this quantitative model, I show how this energy can be channeled away from the qubit so that this error mechanism can be reduced by many orders of magnitude.  I also comment on how this affects other solid-state qubits.  
\end{abstract}
\maketitle

Quantum computers are proposed to perform calculations that cannot be run by classical supercomputers, such as efficient prime factorization or solving how molecules bind using quantum chemistry \cite{shor,qchem}.  Such difficult problems can only be solved by embedding the algorithm in a large quantum computer that is running quantum error correction.  

Quantum computers have intrinsic errors, so algorithms can be natively run with typically only a few hundred to thousand logic operations \cite{nisq, arute}.  In order to run the most powerful and useful algorithms, say with millions to billions of logic gates, errors must be reduced to a parts per million or billion range, or lower.  Fortunately, this is possible using quantum error correction, where the qubit state is distributed to many physical qubits in a way similar to classical error correction, so that errors in the physical qubit states can be selectively measured, decoded and corrected.  For example, surface code error correction encodes a protected “logical” state with about 1000 physical qubits \cite{scorigin,sc}.  As long as physical errors are small, about 0.1\%, and occur randomly and independently among these 1000 qubits, then the logical error can be less than 0.1 part per billion \cite{scaleup}.  However, if errors are large or correlated, bunching together either in time or across the chip in space, then error decoding fails.  With a logical error, the memory of the quantum computer is lost and the algorithm fails.  

This paper explains how cosmic rays and background gamma ray radiation are pulsed energy sources that produce large and correlated errors in superconducting qubits.  Although quasiparticle and radiation effects have been observed in previous experiments on qubits, their \textit{average} effect is small and roughly the magnitude of other decoherence mechanisms ($T_1 \sim 100\,\mu$s) \cite{Serniak,mit}.  On a \textit{per event} basis, however, the effect is argued here to be large ($T_1 \sim 1\,\mu$s) and correlated in area ($\sim\textrm{cm}^2$) and time ($\sim$ ms), which would strongly kill quantum error correction.  It is imperative to further understand this error mechanism, not just reducing the effects of radiation, but slashing the resultant quasiparticle density, time and length scales by a factor of 100 or more \textrm{each}, summarized as an overall million-fold improvement in present designs.  There is no guarantee that such a huge improvement can be extrapolated from present experimental knowledge, so this paper overviews and suggests the critical physics and design changes needed to solve this important bottleneck. 

Cosmic rays naturally occur from high energy particles impinging from space to the atmosphere, where they are converted into muon particles that deeply penetrate all matter on the surface of the earth.  When the muons traverse the quantum chip, they deposit a large amount of energy in the substrate of the quantum processor, on average 460 keV \cite{robert}, which then briefly ``heats'' the chip. Gamma rays from natural background sources have a somewhat larger rate and can deposit energy up to about 1 MeV \cite{robert}.  Experiments on low-temperature detectors and qubits have observed such radiation and quasiparticle effects \cite{robert,oldmicrocal,dirtyal,gransasso,mit,rami,matt,sergeev,devisser,Catelani}.

A model is presented here that quantitatively describes the phonon down-conversion process to quasiparticles, which then decays the qubit state.  This model shows that greater than 90\% of the radiation energy is converted into phonons \cite{cabrera,robert}.  For present designs, 57\% of the phonon energy then breaks Cooper pairs, with significant consequences: these quasiparticles reduce the qubit energy decay time $T_1$ in the range $0.16-1.6\,\mu\textrm{s}$ \cite{nqp,rateqp}, have a large spatial extent of mm to cm range, a long duration of 100\,$\mu$s to 10 ms, and occur more than once per minute.  Each of these parameters is troubling, but in combination they are large enough to kill a complex quantum computation by many orders of magnitude.  

Using a quantitative model for the generation of quasiparticles and their decay, I show that one can reliably redesign the quantum processor by channeling the phonon energy away from the qubits.  The most important change is using \textit{thick} films of a normal metal or low-gap superconductor to channel energy away from qubits.  This redesign should reduce the initial quasiparticle density by a factor of 100, usefully larger than for a previous detector experiment with \textit{thin} films \cite{rami,Valenti,Henriques}.  This work is also complementary to a recent paper that describes well the radiation physics and the effects of breaking electron-hole pairs in the silicon crystal as part of the down-conversion process \cite{robert}; such charge offsets should not be an issue with large transmon qubits \cite{transmon}. 

It is tempting to work on various mitigation strategies to reduce the background radiation, such as using low-radiation materials or running a quantum computer deep underground.   However, as the error rate needs to be reduced to less than one per day, it is unlikely these strategies will be effective enough.  Instead, the energy should be channeled away from the qubits at the chip level.  

In this paper, section 1 reviews the basic conversion physics of radiation to phonons to quasiparticles, how quasiparticles affect the qubit, and how quasiparticles eventually relax by recombination.  In section 2 detailed models are presented that describe an event in 5 stages, predicting the approximate magnitude, time and size scale of the effect.  Section 3 describes the critical stages and how redesign of the qubit can lessen the impact by a huge margin, the goal of this paper.  In section 4 the impact of such radiation is briefly discussed for other qubit systems. 

\section{1. Phonon Down-conversion to Quasiparticles} 

\begin{figure}[t]
\includegraphics[width=0.48\textwidth, 
trim = 50 60 270 80,clip]
{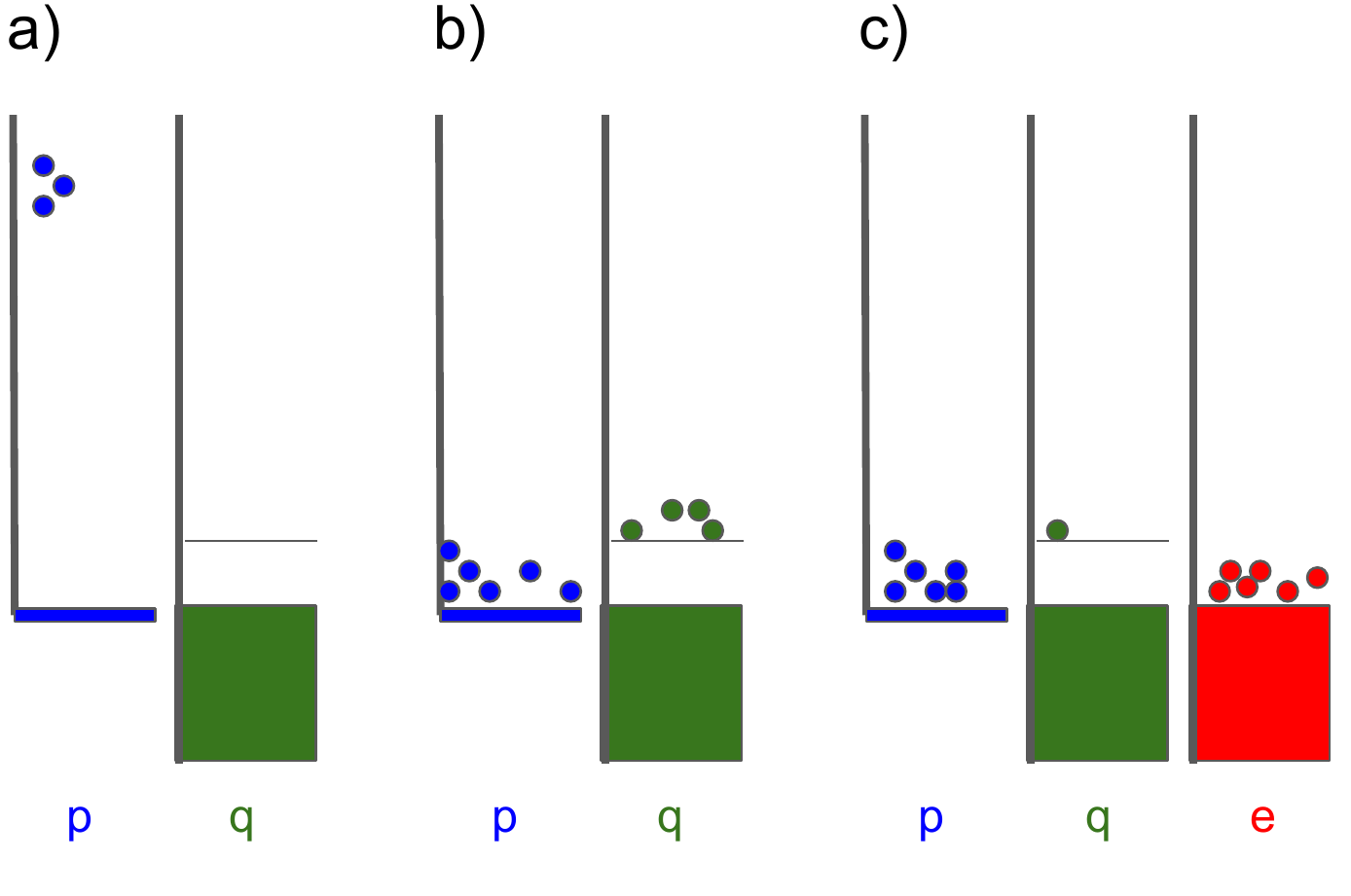}
\caption{\textbf{Non-equilibrium phonon, quasiparticle and electron excitations.}
a) Schematic representation of the phonon (p) and quasiparticle (q) excitations after the radiation event, with only a few high-energy phonons.  The vertical axis represents the state energy.  b) After relaxation, the phonon energy has down-converted to quasiparticle excitations in the superconductor and low-energy (sub-gap) phonons.  About 57\% of the initial energy remains in the quasiparticles.  c) With a normal metal structure, most of the energy has been channeled away from the superconductor and towards the normal metal electrons (e), which is designed not to affect the qubit.  
}
\label{fig:down}
\end{figure}

At low temperatures the physics of thermalization often becomes slow, and sometimes the riskiest assumption is that a system can be simply described by a temperature.  For the non-equilibrium physics described here, it is better to represent the system being mostly at a low background temperature $T$, but with a small number of excitations that can each be described by their energy $E$ or their occupation $f(E)$.  The basic idea is illustrated in Figure\,\ref{fig:down}, where after the radiation event (a) there are a few high-energy phonons, which then get down-converted over time (b) to a larger number of low-energy phonons and quasiparticles. The quasiparticles radiate phonons until they are within typically 50 - 100\,mK of the superconducting gap $\Delta$ \cite{nqp}.  As shown in Fig.\,\ref{fig:down}(c), the excitations need to be channeled to the normal metal and away from the qubit superconductor.  

Following Eqs.\,(3) and (4) of Ref.\,\cite{nqp}, the density of superconducting quasiparticles $n_{qp}$ is
\begin{align}
    \frac{n_{qp}}{n_{cp}} &= \frac{2}{\Delta} \int_\Delta^\infty d\epsilon\  f(\epsilon)\,\rho(\epsilon) \ ,\\
    \rho(\epsilon)&=\epsilon/\sqrt{\epsilon^2-\Delta^2} \ ,
\end{align}
where $n_{cp} = 2.8\cdot10^{6}/\mu\textrm{m}^3$ is the density of Cooper pairs for aluminum and $\rho(\epsilon)$ is the normalized superconducting density of states.  In the two-fluid model of a superconductor, the quasiparticles damp the qubit like a low density normal metal, with a quality factor for a phase or transmon qubit given by \cite{nqp}
\begin{align}
    1/Q \simeq 1.2\ n_{qp}/n_{cp} \, \label{eq:Q}
\end{align}
where the numerical factor 1.2 is computed for 5\,GHz.  As discussed in the appendix, this uses the regular assumption that quasiparticle energies are near the gap.  Here, what matters is the quasiparticle density at the junction, since the dissipation in the two-fluid model can be thought of as coming from the normal-state tunnel resistance $R_T \simeq 100\,\Omega$.  The quasiparticles also provide damping from the superconducting leads of the qubit, but typically can be neglected because its normal resistance is small $\sim 10\,\textrm{m}\Omega \ll R_T$.

The down-conversion of the cosmic or gamma ray radiation has been discussed previously for metals \cite{Kozorezov} and silicon \cite{Martinez}. The initial interaction of the radiation produces silicon electron-hole pairs and high-energy phonons up to the Debye energy 450\,K \cite{robert}.  These phonons down-convert to lower energy phonons, but this process stops at an energy about $50\,$K (1 THz) \cite{Martinez} where the phonon dispersion relationship becomes linear.  In silicon, most of the electron-hole pairs recombine rapidly (ns time scales), with less than 10\% of the radiation energy contributing to their net occupation \cite{cabrera,robert}.  A sapphire substrate should be similar to silicon because both are ``high-gap'' insulators at cryogenic temperatures, but this assumption should be experimentally checked.  

The most important part of the down-conversion process is for phonon energies below 50\,K and the subsequent cascade of phonon and electron/quasiparticle scattering processes in metal.  Here, a high-energy phonon is converted to electron/quasiparticle pairs, which then scatters again to produce another lower-energy phonon.  The relevant scattering processes are listed in Table\,\ref{tab:scatt}.  These down-conversion processes are described by Kaplan \ea\ for superconductors \cite{kaplan}, but can also be used for normal metals by setting the gap to zero.  As these rates are written as integrals over the Fermi occupation factors, the formulas can thus be rewritten to express a single-particle scattering rate versus energy. Rates at $T=0$ but for non-equilibrium quasipartices are given below.   

\begin{table}[t]
\begin{tabular}{| c | c | c |}
\hline
\ \ scattering \ \ & \ Kaplan (Al) \ &  \\
\hline
p $\rightarrow$ q + q & (1/1.0\,ns)($E_p$/K) & \\
q $\rightarrow$ q + p & (1/1700\,ns)$((E_q-\Delta)$/K)$^3$ &  \\
\hline
q + q $\rightarrow$ p & (22/440\,ns) $n_{qp}/n_{cp}$ &  \\
\hline
\hline
\ \ scattering \ \ & \ Power (n-Al) \ & 
\ \ \ Power (Cu) \ \ \ \\
\hline
p $\rightarrow$ e + e & (1/3.1\,ns)($T_p$/K) 
& (1/8.2\,ns)($T_p$/K) \\
e $\rightarrow$ e + p & (1/350\,ns)($T_e$/K)$^3$ 
& (1/24\,ns)($T_e$/K)$^3$ \\
\hline
\end{tabular}
\caption{\textbf{Scattering processes.} Table of scattering processes for phonons (p), electrons (e) and quasiparticles (q), with initial energies $E_p$, $E_e$ and $E_q$ respectively given in Kelvin.  The ``Kaplan'' column represents the scattering rate given by reference \cite{kaplan}, for aluminum.  The ``Power'' column is a rate from the power calculation given in the text, for normal aluminum and copper.  Both give estimates for scattering rates and their energy dependence.  } 
\label{tab:scatt}
\end{table}

\begin{figure}[t]
\includegraphics[width=0.48\textwidth, 
trim = 90 10 190 80, clip]
{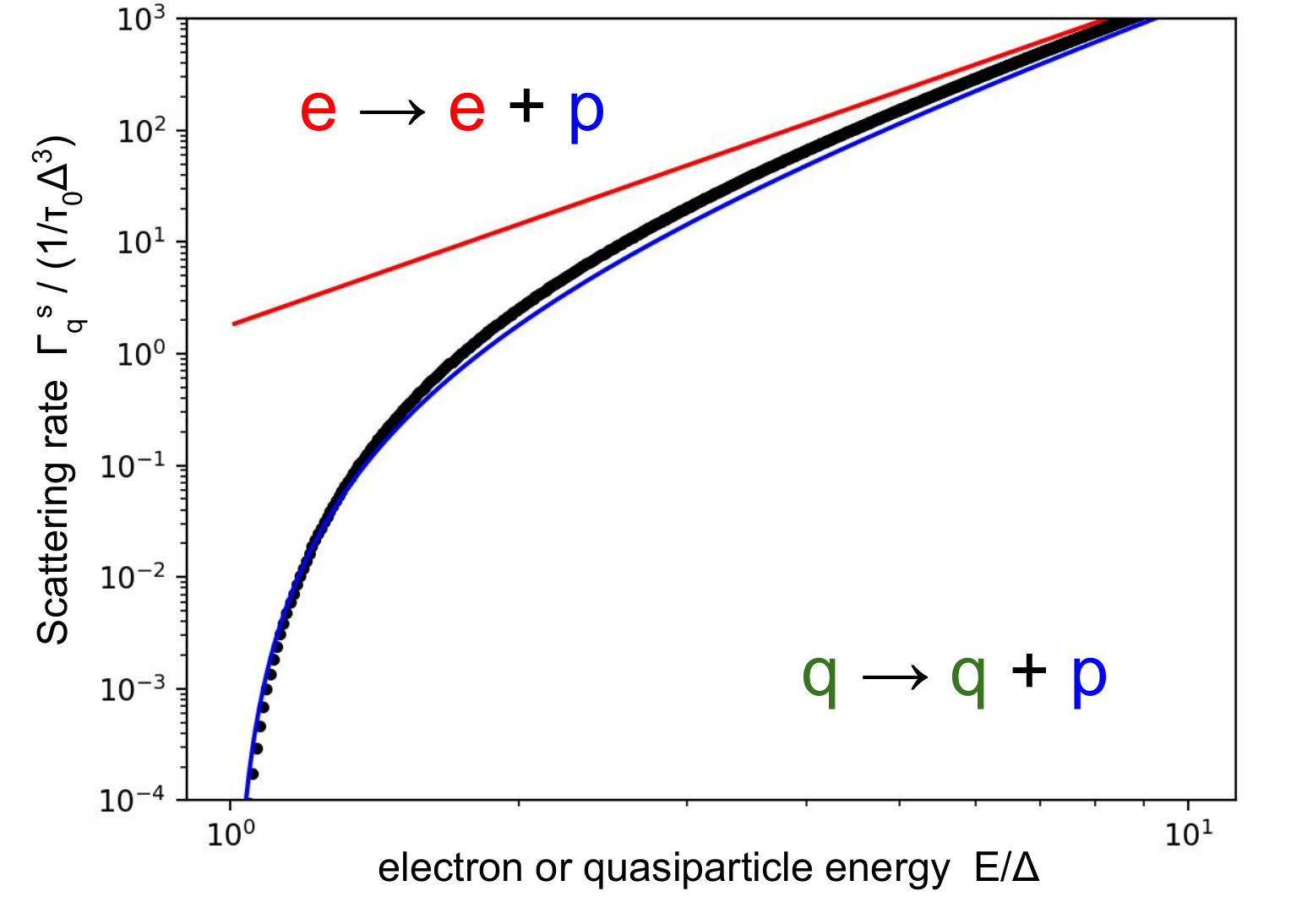}
\caption{
\textbf{Quasiparticle scattering rate.} 
Plot showing the quasiparticle decay rate versus energy (black dots), along with a fit formula $1.8\, (E_q-\Delta)^3$ (blue line); the corresponding rate for a normal metal is the red line.  The rates scale approximately as the final phonon density of states.    
}
\label{fig:QEscatt}
\end{figure}

For $q \rightarrow q + p$, the lifetime of a quasiparticle at energy $\epsilon$ to scatter to any energy $\epsilon’$ by emitting a phonon of energy  $\epsilon-\epsilon’$ is given by Eq.\,(8) of Ref.\,\cite{kaplan}
\begin{align}
    \Gamma_q^s(\epsilon) &= \frac{1}{\tau_0}
    \int_\Delta^\epsilon d\epsilon' \frac{(\epsilon-\epsilon')^2}{(kT_c)^3}
    \rho(\epsilon')
    \Big(1-\frac{\Delta^2}{\epsilon\epsilon'}\Big) \label{eq:qpsc}\\
    & \simeq \frac{1.8}{\tau_0\Delta^3} (\epsilon-\Delta)^3 \label{eq:qpsca}\ ,
\end{align}
where the second equation was fit to the results of numerical integration, plotted in Fig.\,\ref{fig:QEscatt}.  Here the BCS relation is used $\Delta = 1.76\,kT_c$, and $\tau_0 = 440\,$ns is the inferred characteristic electron-phonon time for Al \cite{kaplan}.  Note this rate scales as the available phonon density of states $(\epsilon-\Delta)^3$.  

For $q+q \rightarrow p$, the lifetime of a quasiparticle state with energy $\epsilon$ to recombine with another quasiparticle of any energy $\epsilon'$ is given by Eq.\,(8) of Ref.\,\cite{kaplan}
\begin{align}
    \Gamma_q^r(\epsilon) &= \frac{1}{\tau_0}
    \int_\Delta^\infty d\epsilon' \frac{(\epsilon+\epsilon')^2}{(kT_c)^3}
    f(\epsilon') \,\rho(\epsilon')
    \Big(1+\frac{\Delta^2}{\epsilon\epsilon'}\Big) \label{eq:qsc}\\
    &\simeq \frac{22}{\tau_0}\frac{n_{qp}}{n_{cp}} \ ,
\end{align}
where for the second equation the quasiparticles are assumed to be near the gap \cite{nqp}.  Note the similarity of this recombination rate to scattering, but here the energy of the emitted phonon $\epsilon+\epsilon'$ is greater than $2\Delta$. With the term $f(\epsilon')$, this rate is proportional to the density of quasiparticles, implying that the recombination rate slows down once the density of quasiparticles becomes small.

\begin{figure}[t]
\includegraphics[width=0.48\textwidth, 
trim = 90 10 210 80, clip]
{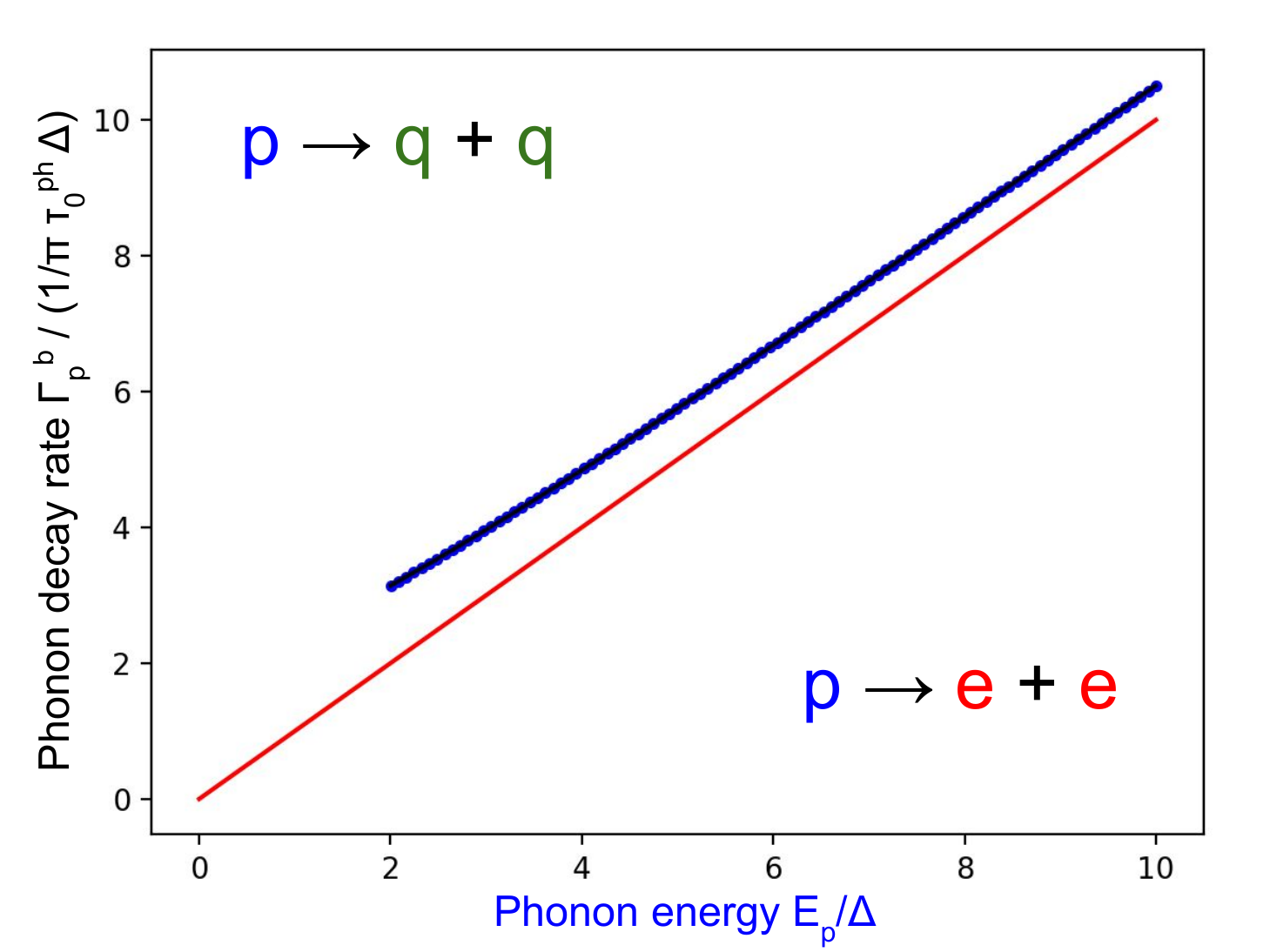}
\caption{\textbf{Phonon decay rate.} 
Plot showing the phonon decay rate versus phonon energy (black dots), compared to a normal metal (red line).  No pair-breaking occurs below energy $2\Delta$, as expected.   The rate above $2\Delta$ is described well by $E_p + 3.8\Delta/(E_p/\Delta+2.5)^{0.8}$ (blue line).
}
\label{fig:Pscatt}
\end{figure}

For $p \rightarrow q + q$, the lifetime of a phonon with energy $E_p$ to break two quasiparticles of energy $\epsilon$ and $E_p-\epsilon$ is given by
Eqs.\,(27) and (30) of Ref.\,\cite{kaplan}
\begin{align}
    &\Gamma_p^b(E_p) = \frac{1}{\pi\tau_0^{ph}\Delta} \nonumber \\
    &\times 
    \int_\Delta^{E_p-\Delta} d\epsilon\,
    \rho(\epsilon)\rho(E_p-\epsilon)
    \Big( 1+\frac{\Delta^2}{\epsilon(E_p-\epsilon)}\Big) \label{eq:psc}\\
    &\simeq \frac{1}{\pi\tau_0^{ph}\Delta}
    \Big[E_p + \frac{3.8\,\Delta}{(E_p/\Delta+2.3)^{0.8}}\Big] \label{eq:psca}\\
    &\sim \frac{1}{\pi\tau_0^{ph}\Delta}\,1.4\,E_p \ ,
\end{align}
where $E_p > 2\Delta$, and the second equation was fit to numerical integration. For Al, Ref.\,\cite{kaplan} gives $\tau_0^{ph} = 0.24\,$ns.   Figure\,\ref{fig:Pscatt} is a plot of the decay rate Eq.\,(\ref{eq:psc}) (black dots) and the excellent approximation of Eq.\,(\ref{eq:psca}) (blue line), and also compares to electron scattering without a gap (red line).  

The phonon energy $E_p$ must be greater than $2\Delta$ to produce two quasiparticles; the excess energy is split randomly between them, on average $E_q = \Delta + (E_p-2\Delta)/2$.  Similarly for quasiparticle scattering, the excess quasiparticle energy is distributed between the quasiparticle and phonon; because the phonon density of states scales as the square of frequency, on average the resulting phonon takes most of the energy $(3/4)(E_q-\Delta)$.  The last scattering process in Table\,\ref{tab:scatt} is recombination, but since it scales as quasiparticle density the rate is small and can be ignored during the initial down-conversion cascade.  

Electron-electron scattering is thought to be unimportant as its rate ($\sim E_e$) becomes relatively small compared to phonon scattering ($\sim E_e^3$) at energies above a few Kelvin.  Also, retaining more energy in the electrons will not change significantly the results here. 

Table\,\ref{tab:scatt} also lists scattering rates based on the power transfer between electrons and phonons, which has been directly measured in experiments.  The power exchange between electrons and phonons within a volume $V$ is given by \cite{roukes}
\begin{align}
    P_{ep} = \Sigma_{ep}\, V\, (T_e^5-T_p^5), 
\end{align}
which looks like the Stefan-Boltzmann radiation law, except here it describes the volume radiation of phonons.   The materials constant $\Sigma$ has been measured for elements and alloys Cu \cite{roukes}, Au \cite{epau}, CuAu \cite{epcuau} and NiCr \cite {epnicr} and found to be reasonably constant $\Sigma \simeq 2\,\textrm{nW}/\mu\textrm{m}^3\,\textrm{K}^5$.  For aluminum the value is about 10 times smaller \cite{epal}.  The total energy of the electrons $U_e$ and phonons $U_p$ are calculated by their integrating heat capacities, with low-temperature values for normal aluminum and copper listed in Table\,\ref{tab:U}.  Along with $P_{ep}$, approximate scattering rates are  
\begin{align}
    \Gamma_p &= P_{ep}/U_p \ \sim T_p  \\
    \Gamma_e &= P_{ep}/U_e \ \sim T_e^3 \ . 
\end{align}
These scattering rates are listed in Table\,\ref{tab:scatt}.  Note that the energy dependence for both the Kaplan and power formulas are the same.  The Kaplan and power rates are similar, showing these are decent estimates for the time scale of electron-phonon physics.  

\begin{table}[t]
\begin{tabular}{| l | c | c |}
\hline
Energy/Power \ \ & \ n-Al \ &  Cu \\
\hline
$U_p$\ \ \  ($10^{-9} \ \textrm{nJ}/\mu\textrm{m}^3/\textrm{K}^4$) & 
$ 2.5 \cdot (T_p^4/4)$ & 
$ 6.6 \cdot (T_p^4/4)$ \\
$U_e$\ \ \  ($10^{-9} \ \textrm{nJ}/\mu\textrm{m}^3/\textrm{K}^2$) &
$\ \ 140 \cdot (T_e^2/2)$ \ \ & 
$\ \ 97 \cdot (T_e^2/2)$ \ \ \\
\hline
$\Sigma_{ep}$ \ \ \ \ \ \ ($\textrm{nW}/\mu\textrm{m}^3/\textrm{K}^5$) & 
0.2 & 2 \\
\hline
\end{tabular}
\caption{
\textbf{Energy parameters from heat capcities.} 
Table of energies for electrons $U_e$ and phonons $U_p$ taken from low-temperature heat capacities for normal aluminum and copper \cite{nistCv}. The electron-phonon coupling constant $\Sigma_{ep}$ is also listed.
}
\label{tab:U}
\end{table}

From these electron, quasiparticle and phonon rates, a length scale for the cascade process can be estimated using the electron and phonon velocities.  The results are summarized in Table\,\ref{tab:length} for a thin $0.1\,\mu$m aluminum film and a thicker $3\,\mu$m copper film, to be introduced later.  The electrons are assumed to diffuse in the metal films; these length scales are typically small compared to phonon lengths.    
  
The most important length is for the phonon interaction, which shows that even at a high energy 20\,K, well above the superconducting gap for aluminum, the interaction length ($0.32\,\mu$m) is somewhat larger than the film thickness ($0.1\,\mu$m).  Thus the phonon in the aluminum will down-convert only a fraction of the time.  When impinging on the aluminum-silicon interface, the phonons will escape the aluminum film about 50\% of the time, travel through the silicon wafer, and are then eventually reabsorbed after another impingement into the metal.  The rate of scattering lowers as the energy approaches $2\Delta = 4\,\textrm{K}$, so that the phonon on average will diffuse laterally by several thicknesses of the silicon substrate, of order a few mm's.

\begin{table}[b]
\begin{tabular}{| l || c | c |}
\hline
Energy      &  20\,K. & 4\,K\\
Material    & Cu/Al & Cu/Al:\\
\hline
e velocity (mm/ns)    & \multicolumn{2}{c}{1.57/2.03} \vline \\
p velocity ($\mu$/ns) & \multicolumn{2}{c}{4.8/6.4}   \vline \\
\hline 
e/q rate (1/ns) & 330/3.5 & 2.7/0.0052 \\
e/q diffuse ($\mu$m)\ \ [for 3/0.1\,$\mu$m]& \ \ \ \ 3.8/7.6\ \ \ \ 
& \ \ \ \ 42/200 \ \ \ \ \\
\hline
p rate (1/ns) & 2.4/20 & 0.49/4.0 \\
p length ($\mu$m) & 2.0/0.32 & 9.8/1.6\\
\hline
\end{tabular}
\caption{\textbf{Scattering lengths.} Table of electron, quasiparticle and phonon scattering rates and lengths for an energy 20\,K, well above the superconducting gap $\Delta$ of aluminum, and for 4\,K $\simeq 2\Delta$ at phonon freeze-out.  The diffusion length scales are for a copper and aluminum thickness of 3\,$\mu$m and 0.1\,$\mu$m, respectively. These rates are only estimates and should be directly measured for proper predictions.} 
\label{tab:length}
\end{table}

\section{2. Present qubit design}

Superconducting qubits are typically fabricated with superconducting metal fabricated on the surface of a silicon chip.  To first calculate the efficiency of high-energy phonons being down-converted to quasiparticles, note that the scattering rates are not important.  This is because high-energy phonons ($E_p > 2 \Delta$) have to eventually scatter in the superconductor, and quasiparticles eventually have to scatter in the superconductor.

\begin{figure}[b]
\includegraphics[width=0.48\textwidth, 
trim = 40 50 300 60,clip]
{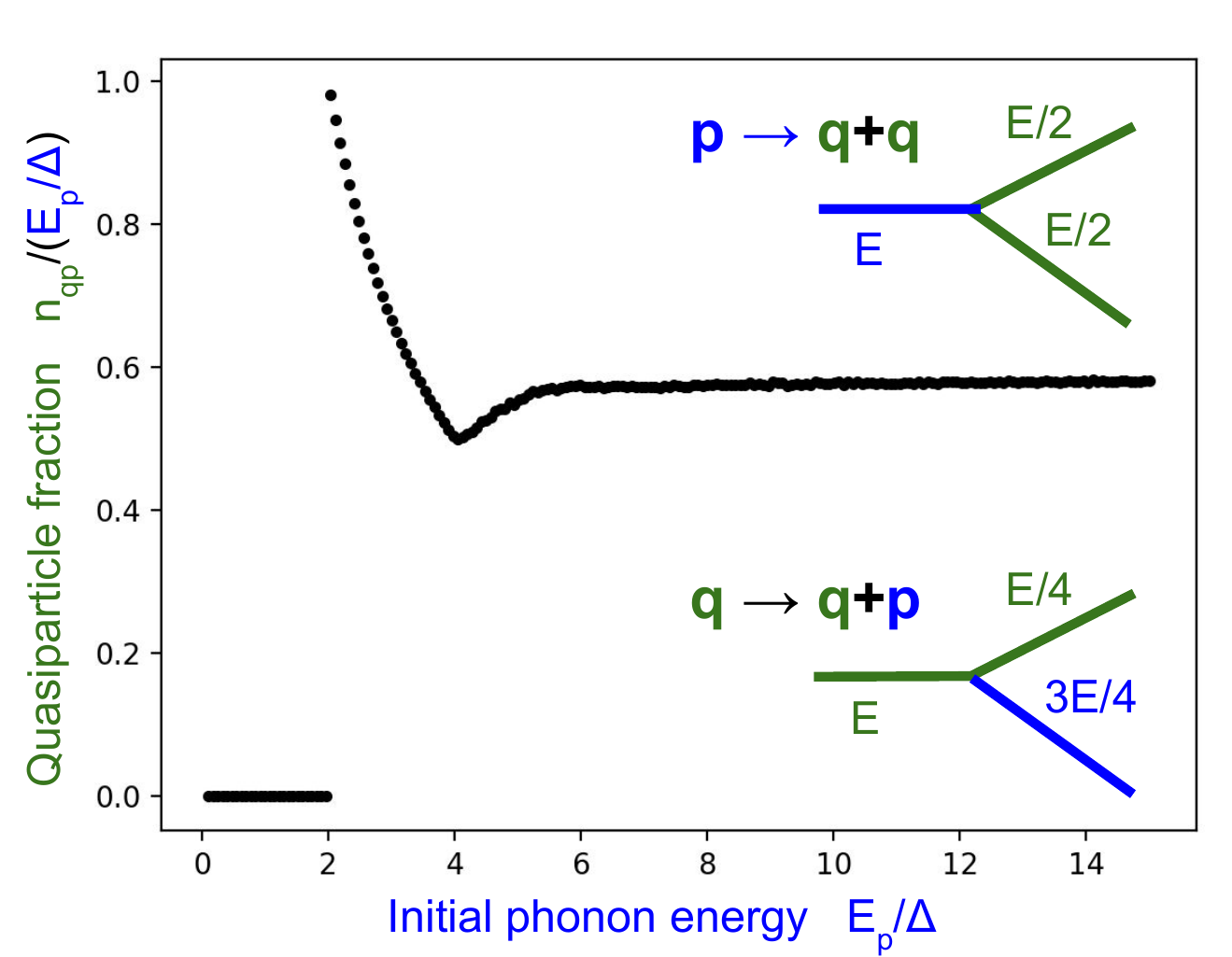}
\caption{\textbf{Phonon to quasiparticle down-conversion efficiency.}
Plot of normalized number of quasiparticles versus initial phonon energy, solved via numerical simulation for the random energy branching of the scattering processes p $\rightarrow$ q + q and q $\rightarrow$ q + p.  No quasiparticles are generated for $E_p < 2 \Delta$, and 2 for $2\Delta < E_p < 4 \Delta$, as expected.  At high energies, 57\% of the initial phonon energy is converted to quasiparticles on average.  The insets depict the scattering process, listing the average kinetic energies.    
}
\label{fig:Qfrac}
\end{figure}

The down-conversion process is numerically simulated using a Monte-Carlo method, where the distribution of final state energies are taken from the integrands of Eqs.(\ref{eq:qpsc}) and (\ref{eq:psc}). The results are shown in Fig.\,\ref{fig:Qfrac}.  The quasiparticle down-conversion efficiency $n_{qp}/(E_p/\Delta)$, which is normalized to the maximum possible number $E_p/\Delta$, is plotted versus the initial phonon energy $E_p/\Delta$.  As expected, at low phonon energies $E_p < 2 \Delta$, there is no breaking of Cooper pairs, but from $2\Delta < E_p < 4\Delta$ one quasiparticle pair is created.  The numerical simulation shows that at larger phonon energies the average quasiparticle fraction is $0.57$, as found previously \cite{tin82,mc92,Kozorezov}, which is reasonable since one expects the initial energy to be shared between both phonons and quasiparticles.  

Assuming the qubits are embedded in a superconducting ground plane, typical for integrated circuits, these down-converted quasiparticles are expected to fill some area of the chip.  The calculations for the down-conversion is show in Table\,\ref{tab:list}.  An initial radiation energy is assumed to be 0.2\,MeV, a typical value for muons and gamma rays.  The next row is the number of quasiparticles using the aluminum gap and conversion efficiency 0.57.  The quasiparticle density is next computed using an aluminum film thickness of $0.1\,\mu\textrm{m}$ and a chip area of $1\,\textrm{cm}^2$.  Next is displayed the fraction density to Cooper pairs, and the estimated transmon quality factor $Q$.  The computed $T_1 = 1.6\,\mu\textrm{s}$ from Eq.\,(\ref{eq:Q}) is short enough to significantly damp the qubit.  As discussed previously, the initial hot-spot of the down-conversion is several times the silicon thickness, so an area of 10\,mm$^2$ may be more appropriate, in which case the decay time is severe, 160\,ns. 

\begin{table}
\begin{tabular}{| l | c |}
\hline
Initial radiation energy  & \ \ \ \ \ 0.2 MeV \ \ \ \ \ \\
Number of qusiparticles & 0.67\,$\cdot10^9$ \\
Density for 1\,cm$^2 \times 0.1\,\mu$m \ \ \ \ \  & 67/$\mu$m$^2$ \\
Density $n_{qp}/n_{cp}$ & $2.4\,\cdot 10^{-5}$ \\
Qubit Q & 51\,k \\
$T_1$ for area 1 cm$^2$ & 1.6\,$\mu$s \\
$T_1$ for area 10 mm$^2$ & 160\,ns \\
\hline
\end{tabular}
\caption{\textbf{Quasiparticle down-conversion and qubit decay.} Table of parameters for down-converting radiation energy into quasiparticles, which then damps the qubits.  The aluminum film is assumed to be 0.1\,$\mu$m thick over the entire 1\,cm$^2$ chip.  Also listed is $T_1$ for a smaller area of the quasiparticle hotspot.  } 
\label{tab:list}
\end{table}

With an understanding of the basic scattering physics, the various stages of the energy down-conversion process are described next.  We assume a low background temperature ($\sim 20\,$mK) well below the gap energy.  The approximate time and length scales are included to estimate spatial and time correlations to the qubit error, along with the qubit decay $T_1$ to understand its magnitude.   

\textbf{1.\,Fireball: 10\,ns, 1\,mm.} The energy spectrum of gamma and cosmic rays are given in Fig.\,S2 of Ref.\,\cite{robert}.  Average cosmic and gamma energies are 0.46 and 0.1\,MeV, but range to about 2.5 and 1\,MeV, respectively.  Cosmic rays deposit energy in a line through the chip, with a spatial extent roughly the thickness of the wafer 400\,$\mu$m, whereas gamma ray energy is in a roughly spherical volume of radius 50-100\,$\mu$m. Since these length scales are smaller than for the various down-conversion and diffusion processes, the two types of events can be treated as equivalent, especially since the purpose here is to minimize the effects of radiation, not measure it.  

After a cosmic ray or gamma ray interacts with the substrate, high-energy phonons and silicon electron/holes are created.  Most of the electron/holes recombine, placing greater than 90\% of the initial energy into phonons.  The phonon energy can down-convert by itself until about 50\,K, when the phonon dispersion relation becomes linear.  These phonons will spread through the substrate, moving away from the creation site.  

\textbf{2.\,Freeze out: 300\,ns, 3\,mm, $\mathbf{T_1 \simeq}$ 160\,ns.} The phonons move through the silicon substrate at a velocity of $9.6\,\mu$m/ns.  When impinging the superconducting metal, about 50\% of phonons will transmit into the metal \cite{Martinez}.  In aluminum, the scattering length of 0.1-0.3\,$\mu$m will cause phonons to down-convert with reasonably high probability ($\gtrsim 50\%)$.  Thus the spatial extent of the fireball stage is several times the substrate thickness.

The quasiparticles and phonons continue to generate a cascade of scattering events, lowering the energy of the phonons until it drops below $2\Delta \simeq 4\,$K.  At this time no additional quasiparticles are generated.  About 57\% of the initial energy is converted to quasiparticles.  As shown in Table\,\ref{tab:length}, the electron diffusion distance is small $\sim 10\,\mu$m, so the lateral extent is set by the phonon diffusion as for the fireball stage.  With $\sim 10\,$mm$^2$ quasiparticle area, the corresponding $T_1$ is given in Table\,\ref{tab:list}.

\textbf{3.\,Qp diffusion: 100\,$\mathbf{\mu}$s, 6\,mm, $\mathbf{T_1 \simeq 0.6\,\mu}$s.} The quasiparticles at first have energy somewhat larger than $\Delta$ and continue to shed their extra energy by emitting phonons.  These phonons cannot be reabsorbed since they have energy below $2\Delta$, and thus bounce around the chip.  For an initial quasiparticle energy 1\,K above the gap, Table\,\ref{tab:scatt} gives a decay time of $1.7\,\mu$s, with a velocity slightly below the Fermi value $v_e$.  For a 0.1\,$\mu$m film, the diffusion distance $D_q$ grows as the square-root of time $t$ 
\begin{align}
    D_q &\simeq \sqrt{v_e t \cdot 0.1\,\mu\textrm{m}} \\
    &= \sqrt{t/\mu\textrm{s}}\cdot 0.4\,\textrm{mm} \ .
\end{align}  
Quasiparticle relaxation times scale as the cube of the energy $E$ above the gap.  At $100\,\mu$s, the quasiparticle energy is approximately 0.25\,K above $\Delta$.  With the quasiparticle velocity dropping as $\sqrt{1-(\Delta/E)^2}$ \cite{trap}, they will diffuse a distance $\sim 0.46\,D_q$. 

This expanding quasiparticle area decreases their density and increases the qubit $T_1$ approximately with $t$.  With velocity lowering with relaxation, the diffusion distance is between 2 and 4 mm, giving a resulting quasiparticle radius of about 6\,mm.  Note that quasiparticles in the superconducting island of a differential transmon will not lower its density by diffusion, so the qubit decay rate $1/T_1$ may not lower over time except by recombination.  

\textbf{4.\,Qp recombination and rebreaking: 1\,ms, chip, $\mathbf{T_1 \gtrsim 1.6\,\mu}$s.}  This stage is concurrent with the last stage of quasiparticle diffusion, but contains additional physics of quasiparticles recombining into Cooper pairs.  Being a two-particle process, the rate is proportional to the quasiparticle density as given in Table\,\ref{tab:scatt}, which produces a density that decreases inversely with time as \cite{rateqp}
\begin{align}
    \frac{n_{qp}}{n_{cp}} = \frac{400\,\textrm{ns}/43.6}{t+t_0} \ .
    \label{eq:qpdecay}
\end{align}
The time is $t+t_0 \sim 100\,\mu$s for the density in Table\,\ref{tab:list} with area 10 mm$^2$, giving the time and size scales for the end of the quasiparticle diffusion stage.  Note that density does not decrease exponentially, so recombination is slow and $T_1$ scales as $1/t$.  Assuming constant quasiparticle area, $T_1$ increases to $20\,\mu$s only after 6\,ms.  Such a long time is roughly consistent with 18\,ms reported in Ref.\,\cite{qptrapvortex}.   

Additionally, it is possible that after recombination the resulting phonon has no loss mechanism other that re-breaking another Cooper pair. Thus the total number of quasiparticles can be roughly constant in this stage.  Here the phonons diffuse throughout the chip because of their long $3.2\,\mu$m interaction length in aluminum.  The re-created quasiparticles will have lower density, and the qubit $T_1 = 1.6\,\mu$s is now computed using the area 1 cm$^2$ of the substrate ground plane .  

Because of this effect, one must be careful in discussing the concept of ``quasiparticle recombination time'', since rebreaking physics might artificially increase this time.   It is imperative to describe the design of the chip mount and how phonons escape, as done in Ref.\,\cite{Vool}.     

Quasiparticle recombination is more complicated because of trapping effects \cite{trap}.  The quasiparticle density at the junction dominates qubit loss since the normal resistance of the junction ($\sim 100\,\Omega$) is so much higher than wiring ($\sim 10\,\textrm{m}\Omega)$.  Quasiparticles will flow to and trap at the aluminum junction if its gap is less than the wiring and ground plane.  Materials defects in the superconducting films can produce localized traps.  Aluminum tends not to trap since scattering raises its transition temperature \cite{altcscatter}, whereas defects in niobium tend to lower its gap.  Magnetic vortices also provide sites for lowering the gap for trapping \cite{qptrapvortex}.  

\begin{figure}[t]
\includegraphics[width=0.48\textwidth, 
trim = 90 250 340 80,clip]
{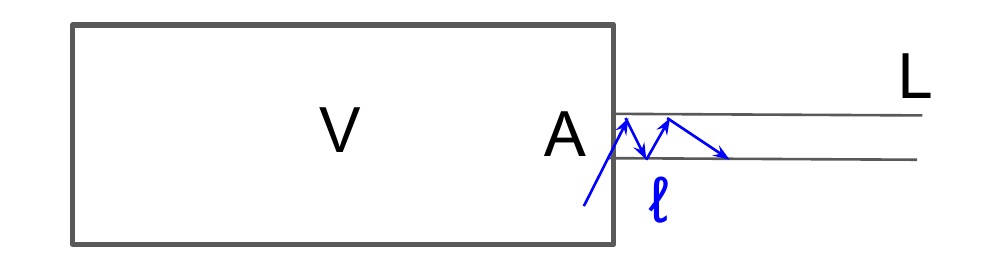}
\caption{\textbf{Device schematic for phonon escape.} Phonons are uniformly moving in a substrate of volume $V$, and can escape through $N_w$ wirebonds each with area $A$.  The escape rate is proportional to the phonon velocity $v_p$.  The phonons must diffuse down the wirebond, eventually escaping with a probability $\ell/L$, where $L$ is the wirebond length and $\ell$ is the mean free path taken as the wirebond diameter.  The escape rate is given by Eq.\,(\ref{eq:Gph}).  
}
\label{fig:escape}
\end{figure}

\textbf{5.\,Phonon escape: 4\,ms, chip.}  The mechanism for the $2\Delta$ phonons to escape is by connection to the thermal ground of the chip mount, which is typically copper.  Our chips are floating to reduce electromagnetic coupling, and only thermalized via wire bonds \cite{bonds}.  The physics for predicting the phonon escape rate is illustrated in Fig.\,\ref{fig:escape}, where the phonons are moving throughout the chip of volume $V$ and can only escape through the $N_w$ wirebonds each of area $A$.  As they must diffuse down the wirebond of length $L$, the exit probability is $\ell/L$, where $\ell$ is the mean free path, here given by the diameter of the wire.  The net phonon escape rate is
\begin{align}
    \Gamma_p &= (N_w A/V) v_p (\ell/L) \ . \label{eq:Gph}
\end{align}
This rate is $\Gamma_p = $ 1/4\,ms for parameters $N_w = 300$, $A=\pi(12\,\mu\textrm{m})^2$, $V=(10\,\textrm{mm})^2(0.4\,\textrm{mm})$,  $\ell = 25\,\mu\textrm{m}$ and $L=2\,\textrm{mm}$.  At time scales much greater than $1/\Gamma_p$ all the phonons, both $2\Delta$ and below, can escape and return the chip to thermal equilibrium.

It has been assumed here that the superconducting gap of the AlSi wirebonds is slightly higher than for the aluminum in the device, so that $2\Delta$ phonons do not break Cooper pairs in the wirebonds.  If they do, the quasiparticles will diffuse in the wirebond with an escape probability that is roughly equal to $\ell/L$, making the above estimate still reasonable.  

\section{3. Improved qubit design}

The first priority when redesigning the qubit chip is to reduce the density of quasiparticles produced in the freeze-out stage.  This can be readily done by adding a normal metal to the substrate so that the phonon energy will be channeled to the normal metal, away from the superconductor.  This normal metal also helps the stage of quasiparticle recombination and re-breaking, channeling the $2\Delta$ phonons to the normal metal instead of back into the superconductor.

The effect of the normal metal can be numerically simulated with a similar method as used for Fig.\,\ref{fig:Qfrac}, except a fraction of the phonon down-conversion is for electron excitations, not just quasiparticles.  Here the fraction is estimated from the relative thicknesses for the superconducting and normal films because the electron-phonon interaction is roughly equal for many metals. Although precise ratios of phonon scattering rates are needed for quantitative designs, this should give a good starting point.

\begin{figure}[b]
\includegraphics[width=0.48\textwidth, 
trim = 100 10 170 70, clip]
{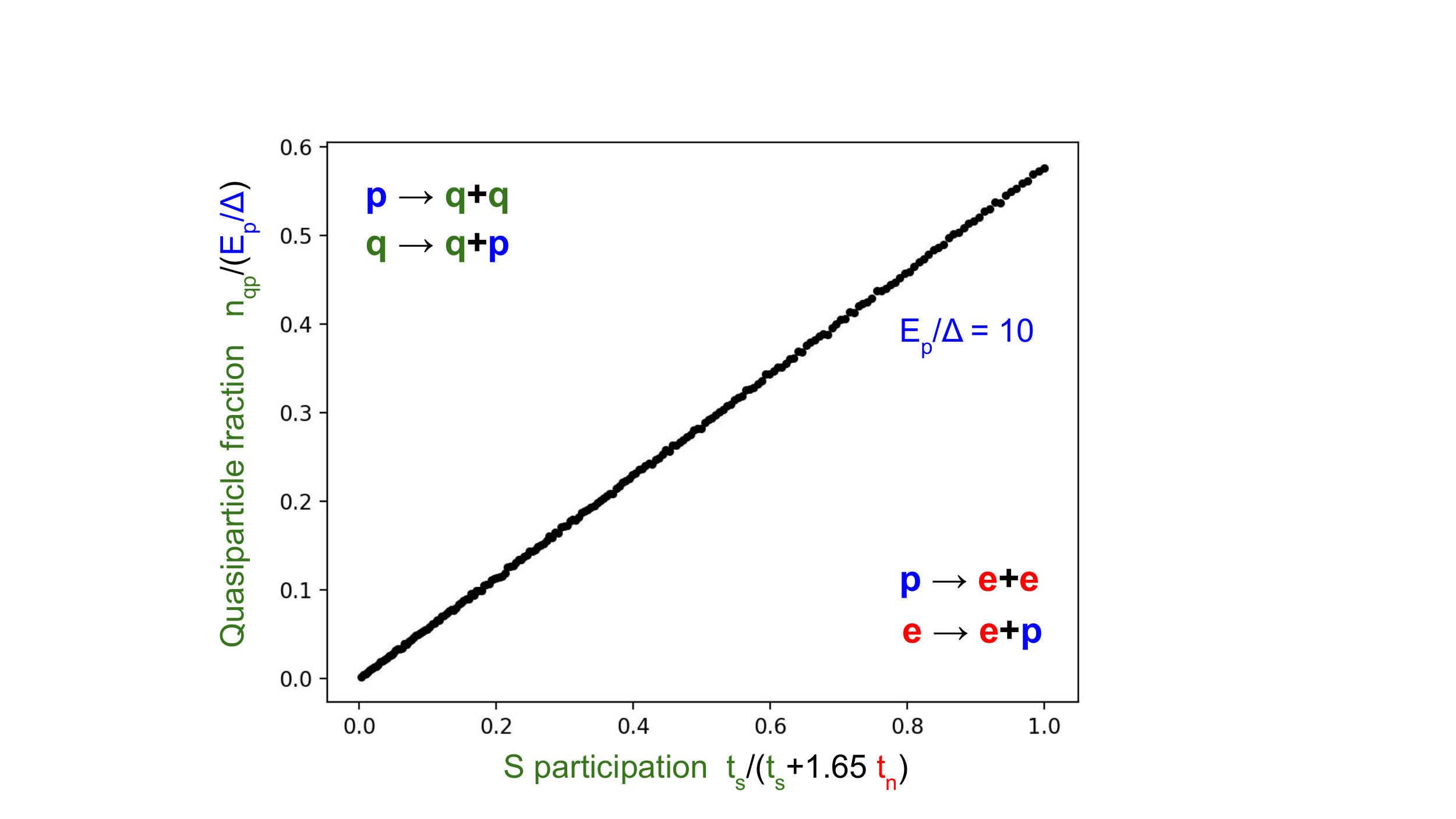}
\caption{\textbf{Reduction of quasiparticles from channeling energy to normal metal.}
Plot of normalized number of quasiparticles versus participation of the superconductor, showing the reduction of quasiparticles using normal metal.  The participation formula is expressed with the thickness of the superconductor $t_s$ and normal $t_n$ metal, fitting the constant 1.65 to give a linear dependence.  The initial phonon energy is $E_p/\Delta =10$.  The insets show the 4 scattering processes for the superconductor and the normal metal.     
}
\label{fig:Qnorm}
\end{figure}

Figure\,\ref{fig:Qnorm} shows the simulation data for the quasiparticle fraction of down-conversion, plotted as the relative thicknesses of the superconducting ($t_s$) and normal ($t_n$) metal.  The participation ratio formula $t_s/(t_s+1.65\,t_n)$ is chosen with the constant 1.65 so that the data lies on a line.  This data shows that as the fraction of superconducting metal decreases, so does its fraction of quasiparticles from phonon down-conversion, as expected.  Prior experiments with thin films ($\sim 0.1\,\mu\textrm{m}$) have shown modest reductions (2-5) of quasiparticles \cite{rami,Valenti,Henriques}. Here, $T_1$ must increase by about a factor of 100 based on estimates from the last section.  Choosing the thickness of the normal metal will ultimately be optimized from experimental measurements of $T_1$, but the above estimates implies a  thickness of about $6\,\mu\textrm{m}$.  As this is significantly thicker than standard thin-film deposition techniques, electro-deposited copper, silver or gold film is likely the best solution.  These metals also have the advantage of low stress. 

\begin{figure}[t]
\includegraphics[width=0.48\textwidth, 
trim = 130 100 190 140, clip]
{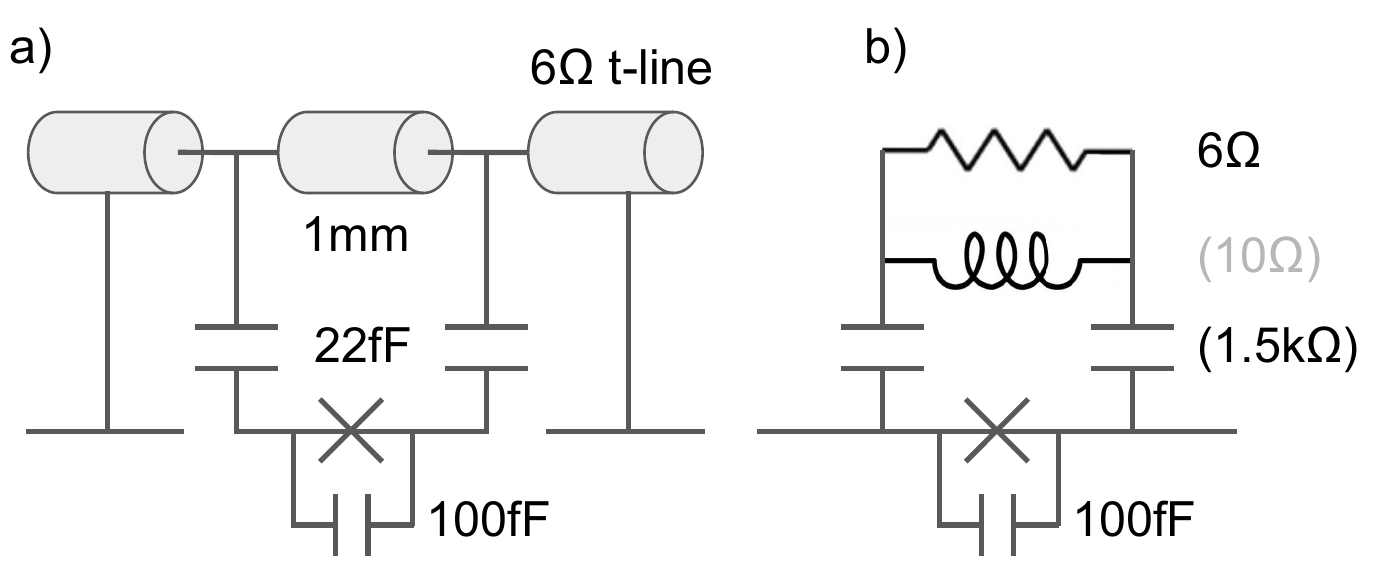}
\caption{\textbf{Equivalent circuit with backside metal.} 
a) Circuit showing differential transmon embedded in a ground plane on the qubit layer.  The continuous normal metal on the backside then acts as a low impedance transmission line of $\sim 6\,\Omega$ impedance, which couples to the transmon via parallel plate capacitors coming from the qubit pads that have an area about 0.1\,mm$^2$.  b) Simplified circuit, replacing the transmission line with an equivalent resistor, and the short transmission line between capacitors with an inductor, with impedances at 5\,GHz shown in parenthesis. This heavily damps the qubit with a $Q \sim 1\,$k.       
}
\label{fig:circ}
\end{figure}

The natural placement of this normal metal film is the backside of the qubit substrate.  However, such a design has problems with electromagnetic radiation.  Figure\,\ref{fig:circ}(a) illustrates the idea with a simplified schematic of the design.  A differential transmon qubit typically is embedded in a ground plane, so that the normal metal on the backside of the substrate forms a transmission line with low impedance, estimated to be $\sim 6\,\Omega$.  The pads of transmon capacitor have an area of about $0.1\,\textrm{mm}^2$ which form parallel plate capacitors that drive the transmission line mode.  Since these capacitors are about 20\% of the capacitance of the transmon, the qubit and transmission line are strongly coupled, so one expects a large effect on the transmon.  This can be estimated by considering the equivalent circuit of Fig.\,\ref{fig:circ}(b), where the short transmission line between the capacitors is equivalently a $\sim 0.3\,\textrm{nH}$ inductor, with an impedance of $10\,\Omega$ at 5\,GHz.  This has a small effect on the equivalent resistance ($6\,\Omega$) representing the transmission line radiation.  The effect of the coupling and this resistor is to significantly damp the qubit with $Q \sim 1\,$k. Note that with strong coupling, this design will not work with any reasonable change to the transmission line impedance. 

A straightforward solution is to break up the normal metal into isolated islands so that the small capacitance between the islands stops the transmission line radiation.  As this only modestly decreases the volume of the normal metal, the thermalization physics will not change much.  For this design the equivalent circuit is given by Fig.\,\ref{fig:circ}(b) but with the resistance given by that across the copper island $\sim 0.01\,\Omega$.  For this low resistance the qubit will be lightly damped with $Q \sim 3\,\textrm{M}$.  If a higher $Q$ is needed, an improvement would be to deposit a thin superconducting film between the substrate and the normal metal, for example titanium.

An alternative design would be to use a continuous superconducting ground plane on the backside, with many via connections to the qubit ground plane to short the transmission line mode.  Then a continuous normal metal film could be used on the backside.  The chip could even placed on a copper mount with vacuum grease to better connect the substrate to thermal ground.  

There is a remaining serious problem: quasiparticles that have been created still have a recombination time that is much longer than the error correction cycle ($\sim 1\,\mu\textrm{s})$, creating errors correlated in time.  This time can be estimated based on Eq.\,(\ref{eq:qpdecay}), which for a qubit $Q = 1\,\textrm{M}$ would require a time to decay $t+t_0 \sim 10\,\textrm{ms}$.  This is clearly too long, even if there is some other mechanism that decays the quasiparticles 10 to 100 times faster.  

\begin{figure}[b]
\includegraphics[width=0.48\textwidth, 
trim = 80 100 240 80, clip]
{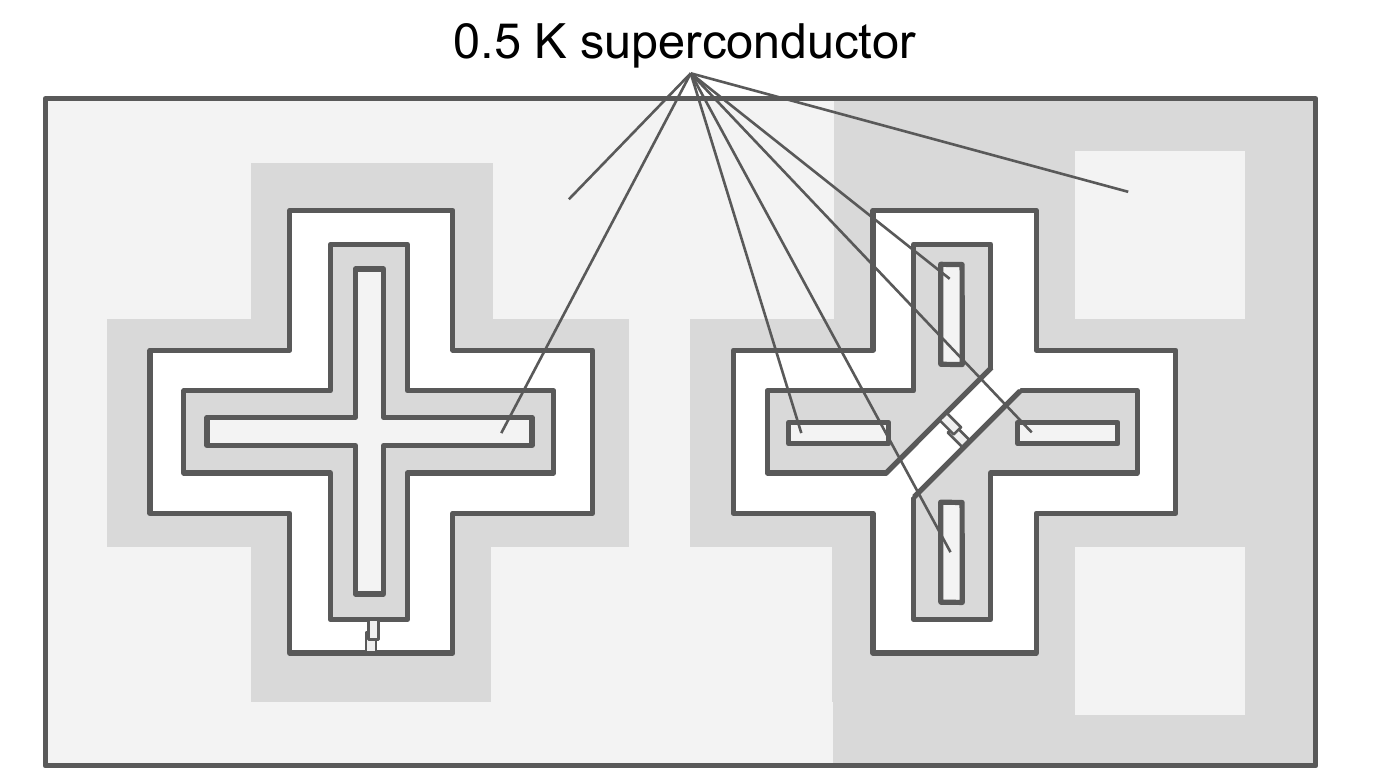}
\caption{\textbf{Design of quasiparticle traps.} 
Illustration of quasiparticle traps for a Xmon (left) and differential (right) transmon, where additional islands in the wiring and ground planes are made from a lower-gap superconductor.  These islands can be directly connected to the wiring or through a thin tunnel barrier.         
}
\label{fig:trap}
\end{figure}

The solution is to build quasiparticle traps into the qubit layer \cite{qptrapping,qptrapnorm,Riwar}, so that any quasiparticles near the Josephson junction can diffuse into a lower-gap superconductor, relax its energy by the emission of a phonon, and thus be trapped away from the junction.  For a transition temperature 0.5\,K, the quasiparticle scattering time is about $1.7\,\mu$s (see Table\,\ref{tab:scatt}), giving a diffusion length of about $0.52\,\mu$m \cite{Riwar}.  With such a length, a significant fraction of quasiparticles might leave the trap before down-conversion, so the design of the trap should have a significant volume fraction.  In order to keep these quasiparticles away from the edges that  are most susceptible to dissipation, a simple solution is to make the islands of the lower gap superconductor, as illustrated in Fig.\,\ref{fig:trap}.  

Reference \cite{Riwar} has a good discussion on the constraints on the resistance of the quasiparticle traps and superconductor interface.  As stated previously for the aluminum leads, qubit loss scales as component resistance: since both the trap and interface resistances can be quite small, about $10^3-10^4$ times less than the junction, their dissipation should be negligible.  For a transition temperature of 0.5\,K, the gap would still be high enough to ensure low dissipation from thermally generated quasiparticles.  Other systems constraints might prevail for the wiring and ground, like choosing superconductors with small surface loss.  

For this improved design, the stages of energy decay should be the following.

\textbf{1.\,Fireball: 10\,ns, 1\,mm.} The initial cascade will have less diffusion laterally because the electron-phonon cascade can down-convert efficiently in the thick normal metal layer.  

\textbf{2.\,Freeze out: 300\,ns, 3\,mm, $\mathbf{T_1 \simeq 16\,\mu\textrm{s}}$.} About 99\% of the phonon energy will be down-converted in the normal metal. Most phonons will not escape this layer even as their energy drops below 1\, K.  Some phonons escaping into the silicon and superconductor may diffuse laterally, but less than for the original design.  Before the quasiparticle energy in the superconductor relaxes to the gap, the quasiparticle density at the junctions is estimated from the energy partitioning.  

\textbf{3.\,Qp diffusion and down-conversion: 1.7\,$\mu$s 3\,mm, $\mathbf{T_1 > 16\,\mu}$s.} The quasiparticles at the junction should diffuse rapidly into the superconductor wiring.  With the distance to the low-gap superconductor islands less than $100\,\mu\textrm{m}$, the diffusion time to the traps is short $< 100\,$ns.  The down-conversion time once in the 0.5\,K superconductor is about $1.7\,\mu$s.  The $T_1$ of the qubit should reset to its background rate during this time.  

\textbf{4.\,Qp recombination: 1\,ms, chip.} During this time, all of the quasiparticles are in the low-gap superconductor, and will recombine over a long time scale as discussed previously.  They should have little effect on the qubit because they are not in the junction, nor in the critical edges of the superconducting wiring or ground plane.  Phonons created by recombination will be absorbed by the normal metal.  

\textbf{5.\,Phonon escape: 4\,ms, chip.}  Phonons continue down-conversion in the normal metal and eventually escape to the copper chip mount.

A summary of the stages and a comparison between the present and future design is shown in Table\,\ref{tab:compare}. 

The qubit lifetime is reduced for a time less than one surface code cycle.  Qubit errors will be small and weakly correlated in space, maybe over one adjacent qubit.  Error correction thus should work properly.

As a side benefit, note that surface loss is presently a dominant damping mechanism, coming from two-level states in surface oxides.  They are likely producing the 0.1-1\% thermal population observed in the qubit excited state, as the two-level states can be excited by non-equilibrium phonons at a frequency $\sim 5\,$GHz.  Strong down-conversion of phonons in the chip should reduce the effective temperature of these or other \cite{Graaf,Serniak} two-level-states and therefore the qubit. Stray quasiparticles are still a concern \cite{Serniak,Houzet,Vool}.  

\begin{table}
\begin{tabular}{| l || c | c | c || c | c | c |}
\hline
\multicolumn{1}{| c ||}{stage}& \multicolumn{3}{c ||}{present} & \multicolumn{3}{c |}{future}  \\
 \hline
  & time & size & $T_1$ & time & size & $T_1$ \\
  & ($\mu$s) & (mm) & ($\mu$s) & ($\mu$s) & (mm) & ($\mu$s) \\
\hline
\hline
1. Fireball & 0.01 & 1 & & 0.01 & 1 & \\
\hline
2. Freeze out & 0.3 & 3 & 0.16 & 0.3 & 3 &  16 \\
\hline
3. Qp diffusion & 100 & 6 & 0.6 & 1.7 & 3 &  $> 16$ \\
\hline
4. Qp recom. \& rebrk. & 1000 & chip & $>$\,1.6 & 1000 & chip & bl \\
\hline
5. Phonon escape  & 4000 & chip &  & 4000 & chip & bl \\
\hline
\end{tabular}
\caption{\textbf{Summary of heating event}.   For the 5 stages, the table summarizes the time scale, size scale, and qubit decay time.  Shown is a comparison of present and improved design. The entry "bl" indicates $T_1$ decay approaches the baseline value. } 
\label{tab:compare}
\end{table}

\section{4. Other Qubit Systems}

Cosmic rays and background gamma radiation is clearly important physics to understand in the systems design of a quantum computer.  Although it has been discussed here in detail for superconducting transmon devices, this physics might be important for other low-temperature qubits.  Here are some comments and questions:

For semiconductor qubits, the charge offsets produced by silicon electron-hole pairs \cite{robert} is as important as the phonon heating discussed here.  This prior work shows that significant charge offsets would be seen over large areas ($100\,\mu\textrm{m}$) for charge-sensitive transmon devices.  Will this introduce correlated qubit errors? 

Photonic quantum computers are being designed based on superconducting transition detectors \cite{photondet}.  Might the phonon pulse described here be enough to trip these detectors in a way correlated in time and space, and thus detrimental to error correction?

Majorana qubits are made from a special quasiparticle state of a super- and semi-conductor.  It would seem that one stray quasiparticle could disrupt this protected state  \cite{Rainis, majorana}.  Is is possible to reduce the quasiparticle number from billions to much less than one so that these states might stay protected?  

More generally, will errors from radiation events disrupt any error-correction scheme that does not use many qubits to protect the state?  Because radiation events produce non-equilibrium energy pulses of size about $100\,\mu\textrm{m}$, are large qubits necessary for errors not to be spatially correlated?  

\section{5. Summary and Conclusion}

Research into thermal particle detectors has enabled low-temperature physicists to understand how radiation injects energy into quantum devices.  The energy pulses from cosmic and gamma rays are clearly an important issue for superconducting qubits, since qubit errors and correlations in time and space will kill error correction by many orders of magnitude.  A model is presented here for this physics, and a solution based on channeling this energy away from the qubit and into benign structures like thick normal metals and quasiparticle traps.  There are many interesting experiments to do soon to demonstrate that an effective solution is possible.

I would like to thank R. McDermott for sharing a preprint before publication, and M. McEwen (a UCSB student working at Google) who shared initial experimental data and patiently waited for this publication. I also thank the referees for thorough reviews.  

\appendix*
\section{Appendix}
\begin{figure}[t]
\includegraphics[width=0.48\textwidth, 
trim = 90 10 190 80, clip]
{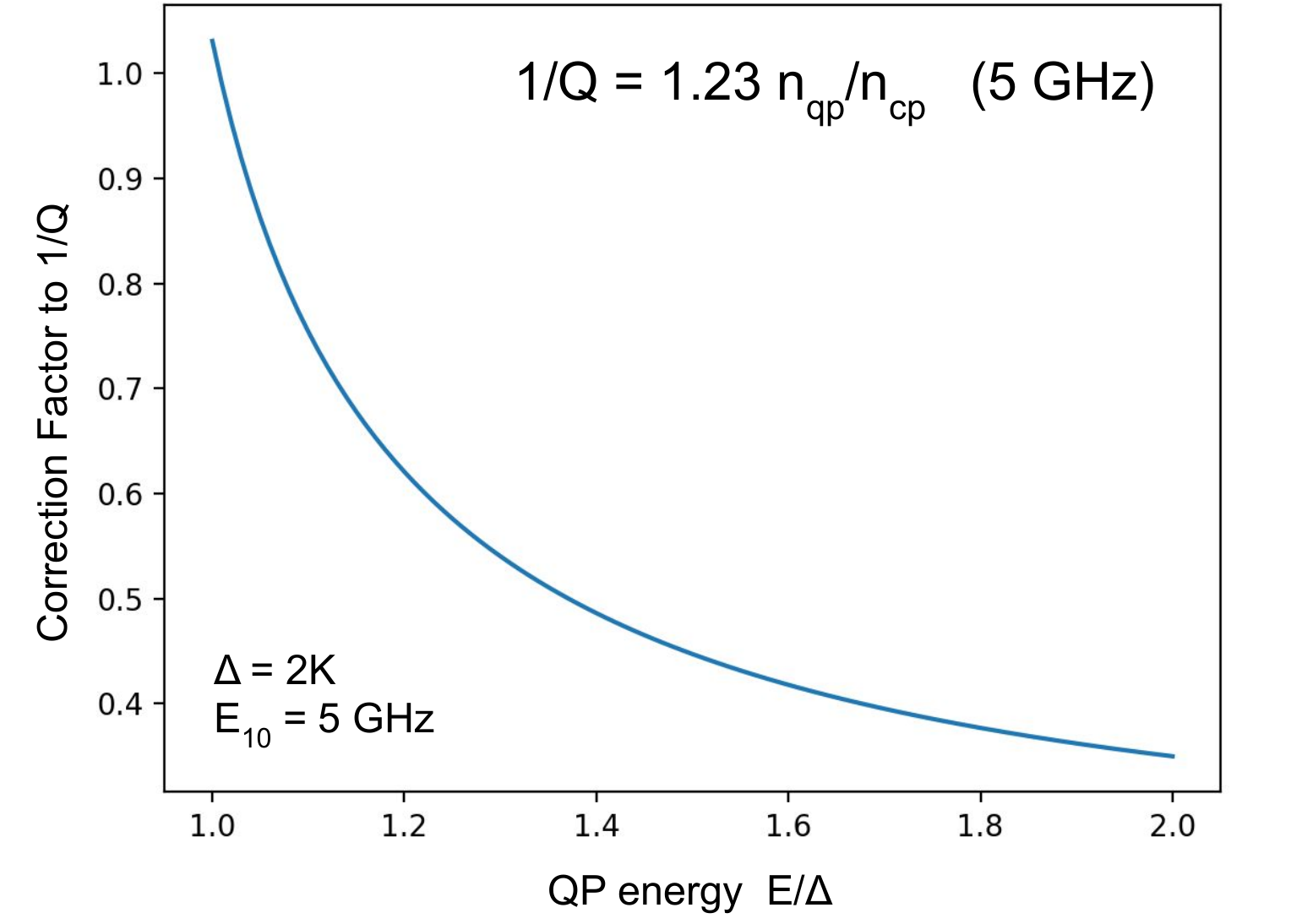}
\caption{\textbf{Qubit decay rate.} 
Plot showing that the qubit decay rate can be lower when quasiparticles have not yet relaxed to the gap energy.  From Eq.\,(3) of Ref.\,\cite{nqp}.
}
\label{fig:qplot}
\end{figure}

Figure\,\ref{fig:qplot} shows the dependence of the qubit energy loss $1/Q$ versus the energy of quasiparticles $E/\Delta$, where all the quasiparticles are assumed to be at energy $E$.  Normally, the quasiparticles are near the gap, but in the freeze-out stage they are somewhat higher in energy, which slightly lowers the qubit decay rate.  

\begin{thebibliography}{10}

\bibitem{shor} 
P. W. Shor, 
Algorithmic Number Theory: First International
Symposium, 
ANTS-I, Lecture Notes in Computer Science 877,
edited by L. M. Adleman and M.-D. A. Huang (Springer,
New York, 1994), p. 289.
\bibitem{qchem}
D. W. Berry, C. Gidney, M. Motta, J. R. McClean, and R. Babbush,
Qubitization of Arbitrary Basis Quantum Chemistry Leveraging Sparsity and Low Rank Factorization, 
Quantum \textbf{3}, 208 (2019).
\bibitem{nisq}
J. Preskill, 
Quantum computing in the NISQ era and beyond,
Quantum \textbf{2}, 79 (2018).
\bibitem{arute}
F. Arute \textit{et. al.}, 
Quantum supremacy using a programmable superconducting processor, 
Nature \textbf{574}, 505 (2019).
\bibitem{scorigin}
S. B. Bravyi and A. Y. Kitaev, arXiv:quant-ph/9811052.
\bibitem{sc}
A. G. Fowler, M. Mariantoni, J. M. Martinis, and A. N. Cleland, 
Surface codes: Towards practical large-scale quantum computation, 
Phys. Rev. \textbf{A 86}, 032324 (2012).
\bibitem{scaleup}
J. M. Martinis,
Qubit metrology for building a fault-tolerant quantum computer,
NPJ Quantum Information \textbf{1}, 15005 (2015).
\bibitem{mit}
A. P. Veps{\"a}l{\"a}inen \ea,
Impact of Ionizing radiation on superconductinng qubit coherence,
Nature \textbf{584}, 551 (2020).
\bibitem{Serniak} 
K. Serniak \ea,
Hot Nonequilibrium Quasiparticles in Transmon Qubits,
Phys. Rev. Lett. \textbf{121}, 157701 (2018).
\bibitem{robert} C. D. Wilen \ea, Correlated Charge Noise and Relaxation Errors in Superconducting Qubits, arXiv:2012.06029
\bibitem{oldmicrocal}
K. D. Irwin, G. C. Hilton, D. A. Wollman, and J. M. Martinis,
X-ray detection using a superconducting transition-edge sensor
microcalorimeter with electrothermal feedback,
Appl. Phys. Lett. \textbf{69}, 1945 (1996).
\bibitem{dirtyal}
Grunhaupt \ea,
Loss Mechanisms and Quasiparticle Dynamics in Superconducting Microwave Resonators Made of Thin-Film Granular Aluminum,
Phys. Rev. Lett. \textbf{121}, 117001 (2018).
\bibitem{rami}
K. Karatsu \ea,
Mitigation of cosmic ray effect on microwave kinetic inductance detector arrays,
Appl. Phys. Lett. \textbf{114}, 032601 (2019).
\bibitem{gransasso}
Cardani \ea,
Reducing the impact of radioactivity on quantum circuits in a deep-underground facility,
arXiv:2005.02286.
\bibitem{matt} 
M. McEwen, 
private communication.
\bibitem{sergeev} 
A. Sergeev and V. Mitin, 
Phonon traps reduce the quasiparticle density in superconducting circuits,
Appl. Phys. Lett. \textbf{80}, 817 (2002)
\bibitem{devisser} 
P. J. De Visser \ea,
Generation-Recombination Noise: The Fundamental Sensitivity Limit for Kinetic Inductance Detectors,
J. Low Temp. Phys. \textbf{167}, 335 (2012).
\bibitem{Catelani}
G. Catelani, R.J. Schoelkopf, M.H. Devoret, and L.I. Glazman,
Relaxation and frequency shifts induced by quasiparticles in superconducting qubits,
Phys. Rev. B \textbf{84}, 064517 (2011).
\bibitem{cabrera}
T. Shutt \ea,
Simultaneous high resolution measurement of phonons and ionization created by particle interactions in a 60 g germanium crystal at 25 mK,
Phys. Rev. Lett. \textbf{69}, 3531 (1992).
\bibitem{nqp}
J. M. Martinis, M. Ansmann, and J. Aumentado, 
Energy Decay in Superconducting Josephson-Junction Qubits from Nonequilibrium Quasiparticle Excitations, 
Phys. Rev. Lett. \textbf{103}, 097002 (2009).
\bibitem{rateqp}
M. Lenander \ea, 
Measurement of energy decay in superconducting qubits from nonequilibrium quasiparticles,
Phys. Rev. \textbf{B84}, 024501 (2011).
\bibitem{Valenti} 
Valenti \ea,
Interplay Between Kinetic Inductance, Nonlinearity, and Quasiparticle Dynamics in Granular Aluminum Microwave Kinetic Inductance Detectors.
Phys. Rev. App. \textbf{11}, 054087 (2019).
\bibitem{Henriques} 
Henriques \ea,
Phonon traps reduce the quasiparticle density in superconducting circuits,
Appl. Phys. Lett. \textbf{115}, 212601 (2019).
\bibitem{transmon}
J. Koch \ea,
Charge-insensitive qubit design derived from the Cooper pair box,
Phys. Rev. A \textbf{76}, 042319 (2007).
\bibitem{Kozorezov} 
A.G. Kozorezov, A.F. Volkov, J.K. Wigmore, A.Peacock, A.Poelaert, and R. den Hartog,
Quasiparticle-phonon downconversion in nonequilibrium superconductors,
Phys. Rev. B \textbf{61}, 11807 (2000).
\bibitem{Martinez} 
M. Martinez, L. Cardani, N. Casali, A. Cruciani, G. Pettinari, and M. Vignati,
Measurements and Simulations of Athermal Phonon Transmission from Silicon Absorbers to Aluminum Sensors,
Phys. Rev. Applied \textbf{11}, 06402 (2019).
\bibitem{kaplan} 
S. B. Kaplan \ea, 
Quasiparticle and phonon lifetimes in superconductors,
Phys. Rev. \textbf{B 14}, 4854 (1976).
\bibitem{roukes}
M. L. Roukes, M. R. Freeman, R. S. Germain, R. C. Richardson, and M. B. Ketchen, 
Hot electrons and energy transport in metals at millikelvin temperatures,
Phys. Rev. Lett. \textbf{55}, 422 (1985).
\bibitem{epau}
D. R. Schmidt, C. S. Yung, and A. N. Cleland,
Temporal measurement of hot-electron relaxation in a phonon-cooled metal island,
Phys. Rev. \textbf{B 69}, 140301 (2004).
\bibitem{epcuau}
F. C. Wellstood, C. Urbina, and J. Clarke,
Hot-electron effects in metals,
Phys. Rev. \textbf{B 49}, 5942 (1994).
\bibitem{epnicr} J. M. Martinis, unpublished
\bibitem{epal}
R. L. Kautz, G. Zimmerli, and J. M. Martinis,
Self-heating in the coulomb-blockade electrometer, 
J. of Appl. Phys. \textbf{73}, 2386 (1993).
\bibitem{nistCv}
At website https://nvlpubs.nist.gov/nistpubs/Legacy/ \\
MONO/nbsmonograph21.pdf
\bibitem{tin82} 
M. Kurakado,
Possibility of high resolution detectors using superconducting tunnel junctions,
Nuclear Instruments and Methods in Physics Research \textbf{196}, 275 (1982).
\bibitem{mc92} 
N. Rando \ea,
The properties of niobium superconducting tunneling junctions as X-ray detectors,
Nuclear Instruments and Methods in Physics Research \textbf{313}, 173 (1992).
\bibitem{trap}
J. Aumentado, M, W. Keller, J. M. Martinis, and M. H. Devoret,
Nonequilibrium Quasiparticles and 2e Periodicity in Single-Cooper-Pair Transistors,
Phys. Rev. Lett. \textbf{92}, 066802 (2004).
\bibitem{qptrapvortex}
C. Wang \ea,
Measurement and control of quasiparticle dynamics in a superconducting qubit,
Nature Comm. \textbf{5}, 5836 (2014).
\bibitem{Vool} 
U. Vool \ea,
Non-Poissonian Quantum Jumps of a Fluxonium Qubit due to Quasiparticle Excitations,
Phys. Rev. Lett. \textbf{113}, 247001 (2014).
\bibitem{altcscatter}
R. W. Cohen and B. Abeles,
Superconductivity in Granular Aluminum films,
Phys. Rev. \textbf{168} 144 (1968).
\bibitem{bonds}
J. Wenner \ea,
Wirebond crosstalk and cavity modes in large chip mounts for superconducting qubits,
Superconductor Science and Technology \textbf{24}, 065001 (2011).
\bibitem{qptrapping}
K.M. Lang, S. Nam, J. Aumentado, C. Urbina, and J.M. Martinis,
Banishing quasiparticles from Josephson-junction qubits: why and how to do it,
IEEE Trans. Appl. Superconductivity \textbf{13} 989 (2003).
\bibitem{qptrapnorm}
R.-P. Riwar \ea, 
Normal-metal quasiparticle traps for superconducting qubits,
Phys. Rev. \textbf{B 94}, 104516 (2016).
\bibitem{Riwar}
R.-P. Riwar and G. Catelani,
Efficient quasiparticle traps with low dissipation through gap engineering,
Phys. Rev. B \textbf{100}, 144514 (2019).
\bibitem{Graaf} 
S. E. de Graaf \ea,
Two-level systems in superconducting quantum devices due to trapped quasiparticles,
Science Advances \textbf{6}, eabc5055 (2020).
\bibitem{Houzet} 
M. Houzet, K. Serniak, G. Catelani, M.H. Devoret, and L.I. Glazman,
Photon-Assisted Charge-Parity Jumps in a Superconducting Qubit,
Phys. Rev. Lett. \textbf{123}, 107704 (2019).
\bibitem{photondet}
A. Verevkin, J. Zhang, and R. Sobolewski,
Detection efficiency of large-active-area NbN single-photon superconducting detectors in the ultraviolet to near-infrared range,
Appl. Phys. Lett. \textbf{80}, 4687 (2002).
\bibitem{Rainis} 
D. Rainis and D. Loss,
Majorana qubit decoherence by quasiparticle poisoning,
Phys. Rev. B \textbf{85}, 174533 (2012).
\bibitem{majorana}
T. Karzig, W. S. Cole, D. I. Pikulin,
Quasiparticle poisoning of Majorana qubits,
arXiv:2004.01264.

\end{thebibliography}
\end{document}